\documentclass[twoside]{ilcws08}
\usepackage[latin1]{inputenc}
\usepackage[dvips]{graphicx,epsfig,color}
\usepackage{wrapfig,rotating}
\usepackage{amssymb,amsmath,array}

\begin{document}
\title{
Asian Test Beam Facilities} 
\author{Satoru Uozumi
\vspace{.3cm}\\
Department of Physics, Kyungpook National University, Daegu 702-701, Korea
}

\maketitle

\begin{abstract}

In Japan, China and Russia, there are several test beam lines available or will become available in near future.
Those are open for users who need electron, muon and charged pion beams with energies of 1-50~GeV for any tests of small-size detectors.
In this manuscript I present a current status of those test beam facilities in the Asian region.
\end{abstract}

\section{Introduction}

There are many detector R\&D activities ongoing for the future linear collider experiment.
Almost all of those activities need to prove those technology by building and testing small test modules or prototypes.
For those tests, good test beam facilities which can provide well-tuned high energy particle beams are indispensable.
Although Fermilab and CERN are quite popular with those excellent test beam sites, some facilities in Asia are also providing useful beams for the small detector tests.

In later sections I briefly introduce following 4 test beam sites in Asia :
\begin{itemize}
\item J-Parc test beam facility,
\item Laboratory of Nuclear Science in Tohoku University,
\item IHEP Test Beam Facility in Beijing,
\item Protovino in IHEP Russia.
\end{itemize}

\section{J-Parc Test Beam Facility}

\begin{figure}[ht]
\centerline{\includegraphics[width=0.9\columnwidth]{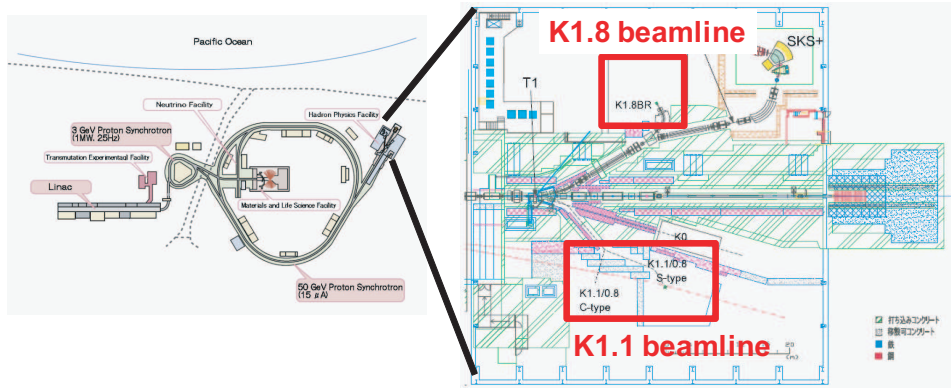}}
\caption{Layout of the J-Parc accelerator and the test beam facility in the target hall.}
\label{Jparc1}
\end{figure}

\begin{figure}[ht]
\centerline{\includegraphics[width=0.9\columnwidth]{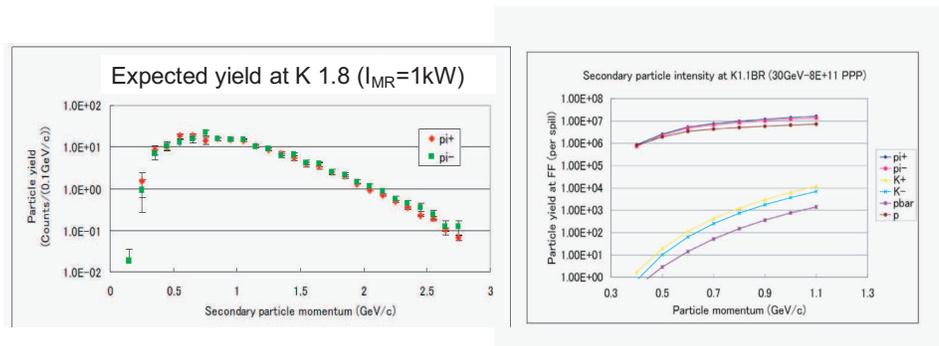}}
\caption{Expected beam rate as a function of beam energy, at the K1.8 and K1.1 beam lines.}
\label{Jparc2}
\end{figure}

The J-Parc accelerator \cite{jparc} in Japan will have the test beam area.
In the target hall, there will be two beam lines named K1.8 and K1.1, as shown in Figure~\ref{Jparc1}, which users can get charged pion, kaon and proton beams up to 3~GeV (on K1.8) or 1.1~GeV (on K1.1).
Originally the K1.8 is the main test beam facility at the J-Parc.
However, during the initial phase of the J-Parc operation, intensity of the main accelerator will not be as much as large, which results in small rate of particles in the test beam area.
Expected rate at K1.8 beam line during the initial operation phase is shown in Figure \ref{Jparc2} left plot.
Due to this reason, the K1.1 beam line is being prepared as the tentative site.
At the K1.1 beam line, even during the initial operation phase users can get high intensity beam as shown in Figure~\ref{Jparc2} right plot.
However the particle energy at K1.1 beam line is limited to 1.1~GeV, and will be available only until the end of 2010.
After the K1.1 becomes unavailable, K1.8 beam line will be opened for users with more intense beams.

\section{Laboratory of Nuclear Science (LNS) in Tohoku University}
\begin{figure}[ht]
\centerline{\includegraphics[width=0.8\columnwidth]{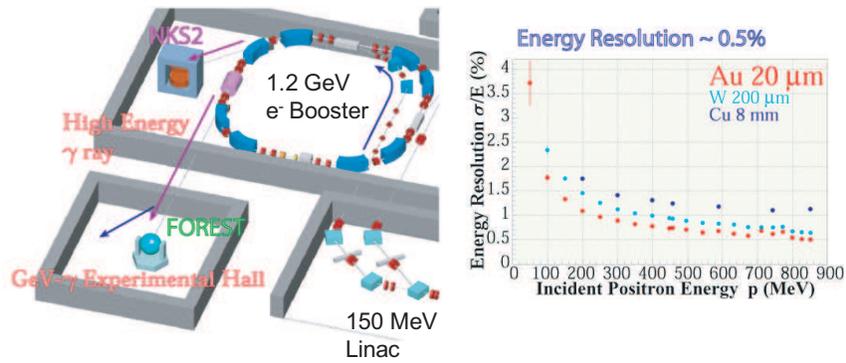}}
\caption{Layout of the Tohoku LNS accelerator facility (right). Users area is located in GeV-$\gamma$ experimental hall in the picture. Left plot is the energy spread of the positron beams for different types of gamma-conversion target.}
\label{tohokuLNS}
\end{figure}

At the Tohoku LNS in Japan, there is an facility where users can get positron beams with energy ranging between0.1 - 0.9 GeV.
It utilizes positrons generated as tartially beam from bremsstrahlung photons by the LNS electron synchrotron.
Layout of the LNS site and energy spread of the electron beam are shown in Figure~\ref{tohokuLNS}.
The beam has 7-seconds on \& 7-seconds off spill structure, and nominal rate during the spill is 2-3~kHz.

\section{IHEP Test Beam Facility in Beijing}
\begin{figure}[ht]
\centerline{\includegraphics[width=0.8\columnwidth]{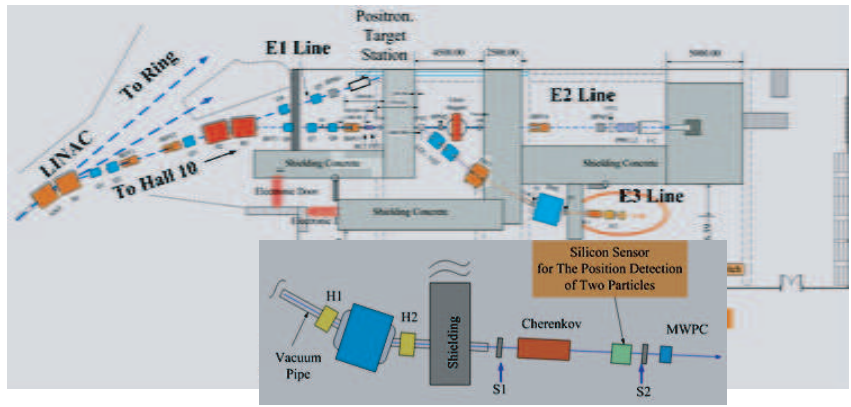}}
\caption{Schematic view of the test beam area in IHEP at Beijing.}
\label{IHEPBeijing1}
\end{figure}

\begin{figure}[ht]
\centerline{\includegraphics[width=0.7\columnwidth]{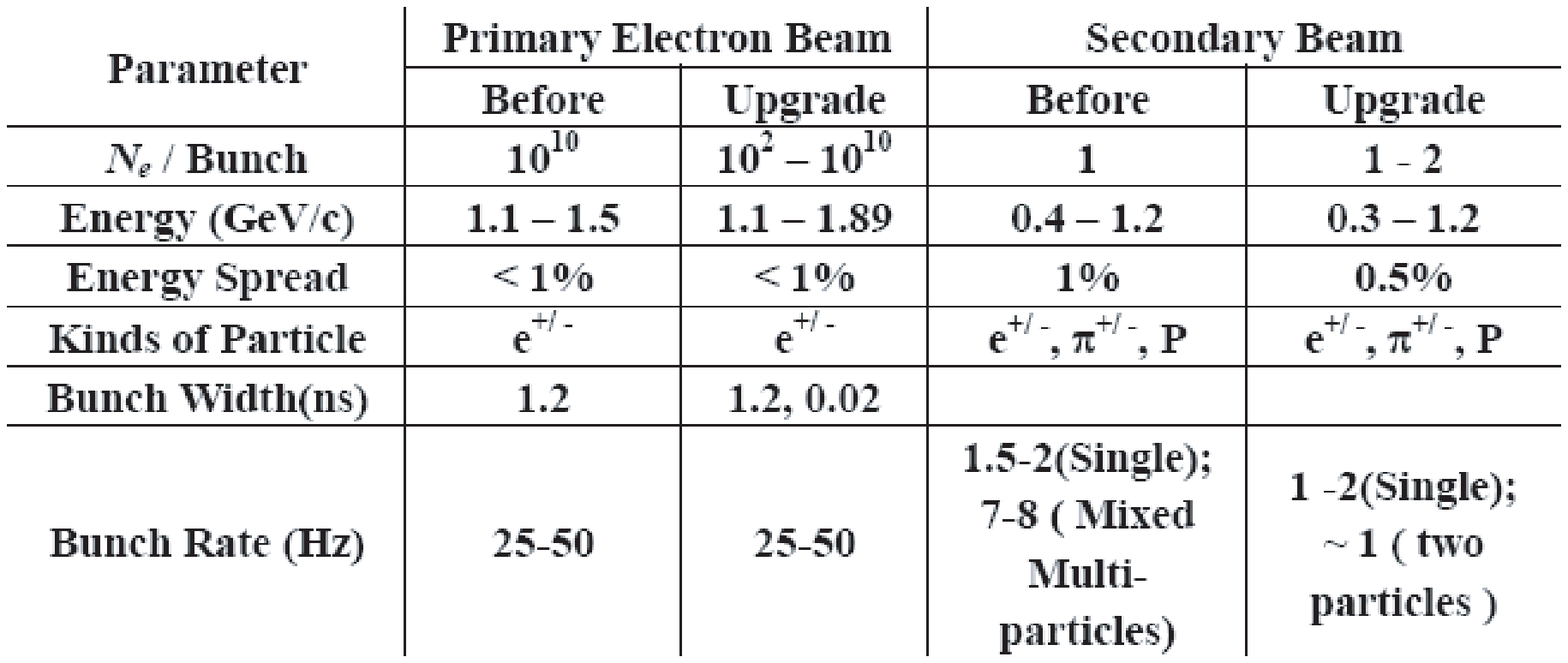}}
\caption{Beam parameters of the IHEP accelerator at Beijing.}
\label{IHEPBeijing2}
\end{figure}

At the end of 2010, the IHEP test beam line at Beijing will be back for users from shutdown period for upgrades.
During the upgrade, several beam line devices, such as Cerenkov detector, silicon tracking device and multi-wire proportional chambers are equipped and available for users use.
Schematic view of the beam line and the beam parameters are shown in Figures~\ref{IHEPBeijing1} and \ref{IHEPBeijing2}.
The beam line will start the commissioning around January 2011, and will provide electron beams up to 1.9 GeV, charged pion and proton beams up to 1.2~GeV to users.

\section{Protovino in IHEP Russia}

\begin{figure}[ht]
\centerline{\includegraphics[width=0.9\columnwidth]{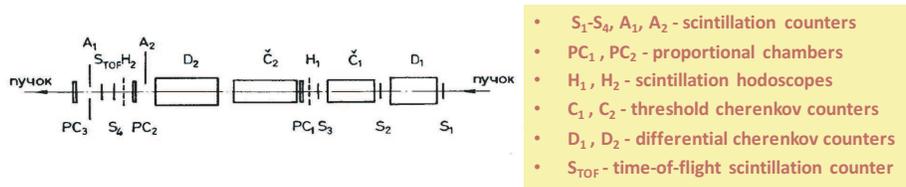}}
\caption{Layout of one of the beam lines (N2B) at IHEP Protovino in Russia.}
\label{IHEPRussia}
\end{figure}

\begin{figure}[ht]
\centerline{\includegraphics[width=0.5\columnwidth]{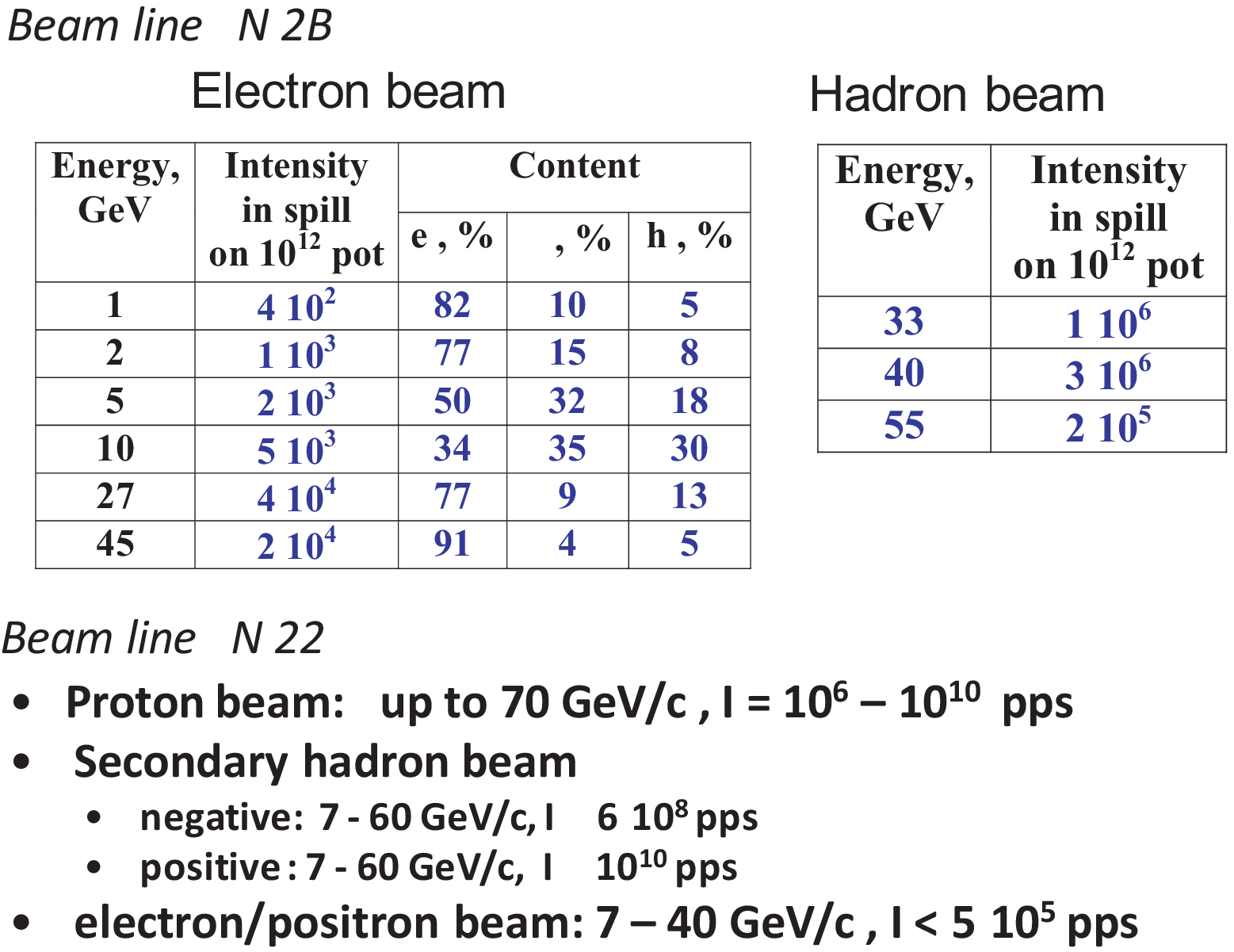}}
\caption{Beam parameters for the test beam facility at IHEP in Russia.}
\label{IHEPRussia2}
\end{figure}

Another test beam facility can be found at Protovino accelerator in IHEP Russia.
Several types of particle beams (electron, charged hadrons) with energy up to 55~GeV are provided to the test beam area.
There are also various beam line devices, scintillation counter, tracker, Cerenkov counter and time-of-flight detectors available for users use as shown in Figure~\ref{IHEPRussia}.
The available periods for test beam users are twice a year, in March-April and Nov-Dec, for a month each time.
During those periods, users can use several beam lines where several types, energies and intensities of particles available.
Table \ref{IHEPRussia2} shows summary of beam parameters for two beam lines N2B and N22.
\section{Summary}
In Asian region there are several beam test facilities available for the linear collider detector R\&D .
Although beam energies and available periods are limited for some sites, those might be still useful for test of small test modules which just needs charged particle beams with order of 1~GeV of energy.


\begin{footnotesize}

\end{footnotesize}


\end{document}